\documentclass[twoside]{article}
\usepackage{qic}
\usepackage{alltt, amsmath,amssymb}

\newtheorem{prop}{Proposition}
\def\se#1{Sec.~\ref{#1}}

\def\de#1{Def.~\ref{#1}}

\def\pro#1{Prop.~\ref{#1}}

\def\eq#1{Eq.\eqref{#1}}

\def\ens#1{\{#1\}}
\newcommand{\ket}[1]{\mbox{$|#1\rangle$}}
\newcommand{\gket}{\ket{\psi}}
\newcommand{\0}{\ket{0}}
\newcommand{\1}{\ket{1}}

\newcommand{\ketbra}[2]{\mbox{$|#1\rangle\langle#2|$}}

\newcommand{\proj}[1]{\ketbra{#1}{#1}}

\def\Rar{\Rightarrow}

\def\Lar{\Leftarrow}

\newcommand{\ox}{\otimes}
\newcommand{\bigox}{\bigotimes}
\def\den#1{\cD({#1})}

\newcommand{\cD}{\mathcal{D}}

\newcommand{\cF}{\mathcal{F}}

\newcommand{\cH}{\mathcal{H}}

\newcommand{\cL}{\mathcal{L}}

\newcommand{\cO}{\mathcal{O}}

\newcommand{\perm}{\mathrm{Perm}}
\def\al{\alpha}

\newcommand{\comp}{\mathbb{C}}

\begin{document}
\setlength{\textheight}{8.0truein}    

\runninghead{The Computational Power of the W and GHZ states} 
            {Ellie D'Hondt and Prakash Panangaden}

\normalsize\textlineskip
\thispagestyle{empty}
\setcounter{page}{1}

\copyrightheading{0}{0}{2003}{000--000}

\vspace*{0.88truein}

\alphfootnote

\fpage{1}

\centerline{\bf
The Computational Power of the W and GHZ states}
\vspace*{0.37truein}
\centerline{\footnotesize
Ellie D'Hondt}
\vspace*{0.015truein}
\centerline{\footnotesize\it Department of Mathematics, Vrije Universiteit Brussel}
\baselineskip=10pt
\centerline{\footnotesize\it Pleinlaan 2, 1050 Brussels, Belgium}
\vspace*{10pt}
\centerline{\footnotesize 
Prakash Panangaden}
\vspace*{0.015truein}
\centerline{\footnotesize\it School of Computer Science, McGill University}
\baselineskip=10pt
\centerline{\footnotesize\it 3480 rue University, Suite 318, Montreal, Quebec, H3A 2A7, Canada}
\vspace*{0.225truein}
\publisher{(May 19th)}{(October 5th)}

\vspace*{0.21truein}

\abstracts{
It is well understood that the use of quantum entanglement significantly
enhances the computational power of systems. Much of the attention has
focused on Bell states and their multipartite generalizations. However, in
the multipartite case it is known that there are several inequivalent
classes of states, such as those represented by the W-state and the
GHZ-state. Our main contribution is a demonstration of the special
computational power of these states in the context of paradigmatic problems
from classical distributed computing. Concretely, we show that the W-state
is the \emph{only pure state} that can be used to exactly solve the problem
of leader election in anonymous quantum networks. Similarly we show that
the GHZ-state is the only one that can be used to solve the problem of
distributed consensus when no classical post-processing is
considered. These results generalize to a family of W- and GHZ-like
states. At the heart of the proofs of these impossibility results lie
symmetry arguments.  
}{}{}

\vspace*{10pt}

\vspace*{3pt}
\communicate{to be filled by the Editorial}

\vspace*{1pt}\textlineskip    
\section{Introduction}
\noindent
The use of quantum resources in computational tasks has led to a revolution
in algorithms~\cite{Nielsen00}.  Models of computation have been developed
which serve as the basis for the design of quantum algorithms.  In the
present paper we study paradigmatic tasks -- leader election and
distributed consensus -- from classical distributed
computing~\cite{Lynch96,Tel94} in the presence of quantum resources.  
Traditional algorithms are typically intended to establish an input-output
correspondence; the main interest in quantum algorithms is in the use of
quantum resources to reduce the time complexity of such algorithms.  By
contrast, our problems are about \emph{joint decision making} by a group of
autonomous agents.  It is much closer in spirit to communication protocols.

As with classical distributed algorithms a whole new arena for establishing
impossibility results becomes available.  The notion of ``universality''
commonly used in algorithms is not relevant here because that notion
assumes that one can entangle any two (or more) qubits.  In a distributed
system one has resources in separated locations and one's actions are
limited to what one can do in a particular region.  One cannot just demand
that a particular global sequence of operations be carried out; it is
necessary to arrange coordination -- usually involving communication --
between the agents acting at the separate locations (regions).  Instead one
has to ask what can be achieved within the limits of a particular
computational model with communication and other limitations built in.  In
our case we study the quantum resources needed to carry out leader election
and distributed consensus in \emph{anonymous} networks.  We focus on
achieving these tasks exactly (to be defined precisely below) rather than
probabilistically.  The ultimate goal is understanding how information
flows between different agents in a quantum setting.

Leader election algorithms have been studied extensively in numerous papers
and for many different network settings.  In networks where processors have
unique identifiers the symmetry is inherently broken via processor
names~\cite{LeLann77}.  By contrast, in an anonymous situation purely
deterministic leader election is impossible: there is nothing that can
break the symmetry if all the processors do the same
thing~\cite{Angluin80}.  If each process has a coin then they can elect a
leader: for example they can each toss a coin and if they get a head they
are the leader.  Of course this is not guaranteed to work, there may be
more than one leader or no leaders and the process will have to be repeated
in the next round.  This idea was first put forward in~\cite{Angluin80} and
later generalized in~\cite{Itai90}.  With probability one this will
terminate eventually but termination is not guaranteed.  There is no bound
on how long this process will take, though the expected number of rounds is
just two.  When two processors share a singlet state they can just measure
it, such that the one who gets, say, one is the leader: this terminates in
one step and always succeeds.  How does this generalize to more than two
processors?  Is such a protocol possible within our framework?

Our main result is that only very special shared quantum resources can be
used to achieve the tasks at hand.  More precisely, we show that in an
anonymous network if the processors share the so-called W-state then a
trivial protocol allows them to solve the leader election problem and, more
importantly, \emph{this is the only possible shared pure state that allows one
to solve this problem at all}.  Of course, if the processors are
allowed to share \emph{mixed states}, or even correlated classical random
variables, then the leader election problem can also be solved.  For example, one can imagine a setting in which all the parties are given a sealed envelope inside which it says if the
party in possession of the envelope is the leader or not. If the
correlations are set up so that only one leader is chosen, having each party opening its envelope elects a leader.  This is essentially the way in which the W-state solves leader election too.  However, we stress that the significant point is not that one can
solve leader election with the W-state: what matters is that one
\emph{cannot} solve it with any other pure state.  We note that our results differ from the work in~\cite{Tani05}, since there quantum communication is allowed to take place between parties. For completeness we also mention~\cite{Pal03}, which treats leader election for non-anonymous networks, and in particular aims at comparing a classical algorithm relying on randomization with its quantum counterpart.

Distributed consensus is usually studied in a fault-tolerant context.  In an asynchronous setting, there is no deterministic algorithm that solves distributed consensus even if only one processor crashes~\cite{Fischer85}. Processor failures can be detected  in synchronous settings, so in this case Byzantine situations are considered, i.e. processors can be malicious. Several algorithms exist in this context, all tolerating up to one third of the processors being malicious. In a quantum setting, we find a trivial protocol which solves distributed consensus with no bound on the number of Byzantine processors. Moreover, by similar techniques as for leader election we find that when ignoring classical post-processing, the GHZ-state is the only shared pure state that allows 
solution of distributed consensus.  These are then essentially
impossibility results: without these states a particular problem is not
solvable in the sense that we make precise below.  The proof uses the
concept of \emph{symmetric configurations}, and as such is similar to the
work of Angluin.  She writes that for anonymous networks `` [\ldots] it
seems intuitively that the behavior of the network can only depend upon
the `local appearance' of the underlying graph''.  Our results are in the
same vein, but with the symmetry breaking being dictated by the (non-local)
properties of the shared entangled quantum state.  The details of the proof
are quite different, however.

The structure of this paper is as follows.  In Sec.~\ref{model} we define
the particular framework in which our results are situated.
Sec.~\ref{moves} formalizes symmetry concepts within this setting, which
are then used in the impossibility proofs of Secs.~\ref{qlesec} and
\ref{qdcsec} concerning leader election and distributed consensus
respectively. We conclude in Sec.~\ref{conclusion}.

\section{The model}\label{model}
\noindent
A distributed system is a system in which several inter-communicating
parties carry out computations concurrently.  While this definition is
quite general, there are a whole series of specific implementations,
depending on the communication structure, the degree of synchronization and
the computational capabilities of individual processors.  The setting of
this paper is that of synchronous, anonymous, distributed quantum networks
with broadcasts, where the network size is known.  Let us clarify these
terms.  First of all, while each processor is allowed a quantum state as
well as a classical state, the communication between processors remains
classical\footnote{Initial qubits are distributed at the time the system is
  set up.}.  Classical and quantum states are assumed to be finite -- in
fact, throughout this paper we work with qubit states.  Communication
occurs via \emph{broadcasts}, which means that all messages are sent out
publicly to all processors along classical channels.  Processors as well as
communication are assumed to be non-faulty.  So far this framework is very
similar to what is known in as LOCC, for \emph{local operations and
classical communication}.

In the anonymous setting, all processors are completely identical, that is
they do not carry individual names with which they can be identified.  As
such the initial network specification must be invariant under permutations
of processors.  One of the implications thereof is that processors start
out in identical local classical states.  However, one has to carefully
restate this when quantum states are allowed.  Indeed, due to the
phenomenon of quantum entanglement, the network quantum state is in general
not a product state of individual (i.e.  local) processor states.  
In this situation, the most sensible definition is to demand that the network quantum state is invariant under permutations of processor subspaces.  This has as an immediate consequence that each processor has the same local view on its quantum state,  i.e. the same reduced density matrix.  Hence, we propose the following definition.

\begin{definition}
An \emph{anonymous distributed quantum network} is such that each processor $p$, where $p \in \{1,\cdots,n\}$ executes the same local protocol and has identical initial classical state, and furthermore that the initial network quantum state $\rho \in \den{\cH_{1}\ox\ldots\ox\cH_{n}}$ is invariant under any permutation of the processor subspaces $\cH_{1},\ldots,\cH_{n}$. In this case we call $\rho$ an \emph{anonymous quantum state}.
 \label{anon}
\end{definition}
Note that anonymity implies that all $\cH_{i}$ are identical. We denote the closure over all permutations of a state $\gket$ by $\perm{\gket}$; for example $\perm{\ket{001}}=\ket{001}+\ket{010}+\ket{100}$. 

Finally, the system is synchronous, which means that a protocol proceeds in a sequence of \emph{rounds}.  During each round a processor receives messages that were sent to it in the previous round,
performs a local operation, and then broadcasts messages.  The local
operation, which in general depends on received messages, consists of
quantum operations followed by a classical post-processing stage, such
that, because of anonymity, each processor has identical decision trees.  

We investigate distributed protocols within this framework, where we make
one further distinction. 

\begin{definition} A distributed protocol is called \emph{terminating} if it reaches a terminal configuration in each
computation, and \emph{partially correct} if the goal of the protocol is achieved for each of the reachable
terminal configurations. A protocol is \emph{totally correct} if it is terminating and partially correct.
\label{tc}
\end{definition}
Note that the above definition does not exclude the possibility of
non-deterministic processes.  In this paper we study totally correct leader
election and distributed consensus protocols.  For a \emph{leader election}
protocol, a correct terminating configuration is one in which there is
exactly one leader in the state \emph{leader}, while all other processes
are in the state \emph{follower}.  In the case of \emph{distributed
  consensus}, all processors should terminate with an identical bit value,
which can be 0 or 1 with equal probability. 
Probabilistic algorithms violate either termination or partial correctness.
Probabilistic leader election algorithms are generally Las Vegas algorithms,
that is, they terminate with positive probability and are
partially correct, i.e.  there are no reachable forbidden configurations.
In the current literature, probabilistic algorithms come about by equipping
each processor with a randomization tool, i.e.  a random number generator
or electronic coin.  In our framework however, we allow only quantum
operations and deterministic classical operations.  Any qubit can of course
be used to implement a coin toss.

These definitions lead to the following propositions.

\begin{prop}
Any totally correct leader election protocol relying on pure states only requires prior shared
entanglement.  
\label{obs}
\end{prop}
\proof {\textit{(outline)}
Without entanglement the network's quantum state is in a product state, and
no entanglement can be created through LOCC operations \cite{Bennett01}.
Therefore each processor is essentially equipped with a coin throughout the
protocol.  As a result, either termination or partial correctness is
compromised, since there is a nonzero probability that local measurement
results are identical}

\begin{prop}
Totally correct leader election algorithms for anonymous quantum networks
are \emph{fair}, i.e.  each processor has equal probability of being
elected leader. 
\label{fair}
\end{prop}
\proof {Recall that the processor state is finite, i.e.  each processor can
  be in 
one of finitely many classical states and has only a finite number of
qubits.  Moreover, with qubits quantum measurements can cause only finite
branching.  Hence there are only finitely many computations, and we can
argue combinatorially.  Suppose then that by a terminating and
correct computation configuration $C$ is reached, wherein, say, process $A$
is elected leader.  Then there exists a corresponding configuration $C'$ in
which process $B$ is elected leader, by defining a permutation of
processors $\sigma$ such that $\sigma(A)=B$ and running $C'$ on the
permuted set of processors.  $C'$ is necessarily correct because of the
anonymity of the network.  This reasoning holds for arbitrary processors
$A$ and $B$ in the network, hence there are equally many terminal
configurations that elect each of the processors as leader respectively,
and our result follows}

The protocols described below are presented with respect to the
computational basis.  Specifically, quantum measurements occur within this
same basis -- this is in fact quite general since measurements in any other
basis can be brought back to these by first applying the appropriate
unitary transform in an anonymous way, i.e.  locally and identical for all
processors.  We frequently denote basis states in their integer
representation if the number of qubits is clear, for example
$\ket{2}=\ket{010}$ for a 3-qubit state.  Normalization factors are
suppressed throughout this paper because the crux of the argument will
depend upon the symmetry rather than on the actual probability amplitudes.
Furthermore, we only address symmetry-breaking capabilities of the quantum
parts of our protocol. For leader election this does not diminish the
strength of our result since it is known from Ref.~\cite{Angluin80} that
classical protocols, and thus classical post-processing, cannot break the
symmetry in anonymous networks.  For distributed consensus a more nuanced
assessment of our result is required.  Unlike the situation with leader
election - there are classical solutions but our quantum protocol is more
fault tolerant.

\section{Symmetric moves}\label{moves}
\noindent
In this section, we prove that certain types of superposition terms are potentially present throughout a
computation. As we shall see below, this situation corresponds to a path of the computation in which a
group of processors evolves symmetrically. We first define the notion of a \emph{k-symmetric move}, which captures particular symmetry properties of a quantum state. We assume that this state is shared between processors, such that each of them owns $m$ qubits.

\begin{definition}
Suppose an $n$-partite state $\gket\in \cH^{\ox n}$, where $\cH$ is a
$2^{m}-$dimensional Hilbert space, is distributed over $n$ processors.  We
say that there exists a \emph{$k$-symmetric move} for the processors
$p_{1}, \ldots, p_{k}$ with respect to $\gket$, where $0 < k \leq n$, if
for all complete sets of orthogonal projectors\footnote{A \emph{complete set of orthogonal projectors} is a set of projectors $\ens{P_{j}, j=1,\dots,J}$ such that $\sum_{j=1}^{J}P_{j}=I$ and $P_{j}P_{j'}=\delta_{jj'}P_{j}$.} $\ens{P_{j}, j=1,\dots,J}$ on $\cH$, with $0<J\leq 2^{m}$, there exist indices $ l, j_{k+1},\dots, j_{n} \in \{1,\ldots,J\}$, with $j_{i}\neq l$ for all $i$, such that
\begin{equation}
(P_{l})^{\ox k}_{p_{1},\ldots,p_{k}}\bigox_{i=k+1}^{n}(P_{j_{i}})_{p_{i}} \gket \neq 0  \text{  .}
\end{equation}
\label{move}
\end{definition}
Note that for any state $\gket$ some incomplete measurements, for which $J<2^{m}$, \emph{always} result in more than $k$ processors obtaining the same results, the trivial example being the identity projector.  The point of \de{move} is of course that it is a statement about \emph{all} possible measurements. In particular, these are by no means restricted to measurements in the computational basis. 

The following lemma concretizes how we can recognize quantum states allowing $k$-symmetric moves.

\begin{lemma}
There exists a $k$-symmetric move for the processors $p_{1}, \dots, p_{k}$ with respect to $\gket\in
\cH^{\ox n}$ if and only if for any basis $\ens{\phi_i} _{i=1}^{2^{m}}$ of $\cH$ there exist indices $ l, j_{k+1},\dots, j_{n} \in \ens{1,\dots 2^{m}}$, with $j_{i}\neq l$ for all $i$, such that, when expressing $\gket$ with respect to the basis $\ens{\phi_i}$, the coefficient of $\ket{\phi_{l}}^{\ox k}_{p_{1}, \dots, p_{k}}\ox_{i=k+1}^{n}\ket{\phi_{j_{i}}}_{p_{i}}$ is nonzero.  
\label{movelemma}
\end{lemma}
\proof {In the proof below we assume, without loss of generality, that  $p_{1}, \dots, p_{k}$ are the first $k$ out of $n$ processors.

\noindent
\boxed{\Rar} For an arbitrary basis $\ens{\phi_i, i=1,\dots 2^{m}}$ of $\cH$ we have that 
\begin{equation}
\gket=\sum_{j_{1}=1}^{2^{m}}\dots\sum_{j_{n}=1}^{2^{m}}\al_{j_{1}\dots j_{n}}\bigox_{i=1}^{n}\ket{\phi_{j_{i}}}\text{  .}
\label{expansion}
\end{equation}
Using \de{move} with $P_{j}=\proj{\phi_{j}}$ for $j=1,\dots,2^{m}$, we find that there exist indices such that
\begin{equation}
(\proj{\phi_{l}})^{\ox k}\bigox_{j=k+1}^{n}(\proj{\phi_{j}})\sum_{j_{1}=1}^{2^{m}}\dots\sum_{j_{n}=1}^{2^{m}}\al_{j_{1}\dots j_{n}}\bigox_{i=1}^{n}\ket{\phi_{j_{i}}}\neq 0\text{  ,}
\end{equation}
or in other words, we find that
\begin{equation}
\al_{l\dots l j_{k+1}\dots j_{n}}\neq 0 \text{  .}
\end{equation}
\boxed{\Lar} For any set $\ens{P_{j}, j=1,\dots,J}$, construct the corresponding basis $\ens{\ket{\phi_i}, i=1,\dots 2^{m}}$ by combining bases for the subspaces described  by each of the $P_{j}$. With respect to this basis there exist indices $ l, j_{k+1},\dots, j_{n} \in \ens{1,\dots 2^{m}}$ such that the coefficient $\al_{l\dots l j_{k+1}\dots j_{n}} $ in \eq{expansion} is nonzero. As a result, we find that $(P_{l})^{\bigox k}\ox_{i=k+1}^{n}(P_{j_{i}}) \gket\neq 0$, and therefore that there exists a $k$-symmetric move with respect to $\gket$}

For example, in a network with three processors, where each processor owns one qubit, the above lemma tells us that there exists a 2-symmetric move with respect to the state $\ket{001}+\ket{100}$ for the first two processors as well as the last two processors, but not for the first and the third processor. Likewise, in a network with two processors where each processor owns two qubits, there is a 2-symmetric move for both processors with respect to the state $\ket{++++}$, but not with respect to the state $\ket{+++-}+\ket{++-+}$. Furthermore, only when $\cH=\comp^{2}$ $k$-symmetric moves imply $(n-k)$-symmetric moves; for example, 1- and $(n-1)$-symmetric moves exist for all network states containing a term of the form $\1{}\0{}^{\ox (n-1)}$.  

Anonymous networks  are by definition invariant under processor permutations, and therefore a $k$-symmetric move, if it exists, automatically exists for arbitrary subsets of $k$ processors in the network. Because of this, we do not specify any particular subset of processors in what follows below. Concretely, in an anonymous setting  the coefficients of all permutations of $\ket{\phi_{l}}^{\ox k}\ox_{i=k+1}^{n}\ket{\phi_{j_{i}}}$ are nonzero by the lemma above.

In our setting, branching during a protocol occurs only due to measurement. If a $k$-symmetric move exists $k$ processors may evolve symmetrically, by following that branch if which they all obtain identical measurement outcomes. In some protocol executions it is possible that subsequent branching continues to conserve the symmetry between $k$ processors in the above way. This leads to the following definition.

\begin{definition}
A protocol has a \emph{$k$-symmetric path} if it there exists a protocol execution consisting of a sequence of $k$-symmetric moves.  
\end{definition}

We then have the following result.

\begin{prop}
Any anonymous distributed quantum protocol for which there exists a $k$-symmetric
move initially has a $k$-symmetric path. 
\label{branch}
\end{prop}
\proof{
Without loss of generality, we assume that local quantum operations during
one round consist of an isometric operation $\mathbf{U}$, i.e.  a unitary
operation along with creation of ancillae, followed by a measurement $M$.
If Def.~\ref{move} holds for the initial network state $\gket$, then it
must hold for $\mathbf{U}^{\ox n}\gket$; indeed, after such an operation we
again obtain a network state as in Lemma~\ref{movelemma}.  Suppose that for
the subsequent measurement $M$ the protocol follows the existing
$k$-symmetric move, corresponding to identical measurement results $j$ and
projections on $\ket{\phi_{j}}$.  Knowing that classical post-processing
cannot break symmetry in anonymous networks, in this case identical
measurement results are broadcast, such that the local operations applied
in the next round, depending on these results, are identical.  Moreover, at
this point the network state still allows a $k$-symmetric move, since it
contains a term of the form $\ket{\phi_{j}}^{\ox k}\ket{\phi_{j}}^{\perp}$.
Then by induction we can construct a $k$-symmetric path for the entire
protocol}

\pro{branch} lies at the basis of the proofs in the rest of this article. It captures the conservation of symmetry properties for specific protocols, which is what we rely on to prove that only particular quantum resources allow an exact solution to the problems of leader election and distributed consensus below.

\section{Quantum leader election}\label{qlesec}

\subsection{One qubit per processor}\label{qleone}
\noindent
In case that each processor owns exactly one qubit of some shared quantum state, we have the following important result.

\begin{theorem}
A necessary and sufficient condition for a totally correct anonymous
quantum leader election (QLE) protocol, where each processor owns 1 qubit
initially, is that processor qubits are entangled in a W-state. 
\label{main}
\end{theorem}
We prove this theorem in both directions below.

\paragraph{W is sufficient.} The idea is to share a specific entangled state between all parties, which
allows one to break the symmetry in one step.  The state used is known as the
W-state, where  

\begin{equation}
W_{n}=\sum_{j=1}^{n}\ket{2^{j}}
\end{equation}
For example $W_{3}=\ket{001}+\ket{010}+\ket{100}$.  This state can be used
as a symmetry-breaking quantum resource.  Each processor $i$ carries out
the following protocol. 

\begin{alltt}
1.  \(q \leftarrow i\)th qubit of \(W\sb{n}\)
    b = 0
    result = wait
2.  b := measure \(q\)
3.  if b = 1 then result := leader
             else  result := follower 
\end{alltt}
This is a totally correct protocol with time complexity $\cO(1)$ and no
message passing.\footnote{We could also have used the state $\overline{W}_{n}$, the complement of $W_{n}$, to carry out the protocol; in this case the processor measuring $\0{}$ becomes the leader.} Note
that the QLE protocol also works for different communication graphs or in
asynchronous networks.

\paragraph{W is necessary.} For this part of the proof we use the tools from \se{moves}.
Specifically, we prove that for certain types of superposition terms in the
initial quantum state total correctness is compromised.  

Suppose $P$ is an arbitrary protocol solving QLE. Then if $P$ allows $k$-symmetric paths with $k$ different from 1 or $n-1$ it is not totally correct.  Indeed, for such a path $P$
either does not terminate or it terminates in a forbidden configuration.
Hence by \pro{branch}, the initial network state $\gket$ cannot allow
$k$-symmetric moves for $k \notin \{1, n-1\}$.  By Lemma~\ref{movelemma} with
$\gket \in (\comp^{2})^{\ox n}$ anonymous in the sense of \de{anon},
this leaves us with $W_{n}$, or unitary transforms thereof, as the only
possibility.

\subsection{More qubits per processor}
\noindent
One can repeat the same symmetry argument in the case where each processor
has $m$ qubits, i.e. $\cH=\comp^{2m}$. Taking $\{\ket{\phi_{i}}\}$ to be a
basis for $\cH$, this leads to initial network states of the form 

\begin{equation}
\gket=\sum_{i=1}^{2^{l}}\alpha_{i}\perm \ket{\phi_{i}}\bigotimes
_{j=2}^{n}\ket{\phi_{i_{j}}} 
\label{genW}
\end{equation}
Here we need one extra ingredient: as before, each processor knows
beforehand which measurement results lead to them becoming a leader.
However, since up to $m$ different results are possible the situation is
slightly more complicated.  So suppose $\cL$ is spanned by the leader
labels $\{\ket{\phi_{i}}, i=1, \ldots, 2^{l}\}$ and $\cF$ is spanned by the
follower labels $\{\ket{\phi_{i_{j}}}\}$.  Then $\gket$ cannot allow
$(k>1)$-symmetrical moves  with respect to $\cL$, which means concretely
that none of the $\ket{\phi_{i}}$ can appear in the tensor product in
\eq{genW}, or in other words we take $\cH=\cL \oplus \cF$.  Note that
\eq{genW} includes the more stringent dual situation where $n-1$
processors are symmetric  with respect to $\cF$.  A QLE protocol would then succeed
by measuring $\gket$, such that the processor obtaining a result in $\cL$
appoints itself leader, while those obtaining results in $\cF$ become
followers.  As a result, we obtain a family of W-like states as the only
possible entanglement resources for totally correct anonymous QLE
protocols.

\subsection{An instructive example}
\noindent
The following example shows the drastic impact of anonymity on the success
of a protocol.  Suppose we have an odd number of processors $n$, and each
processor carries out the following protocol\footnote{$H$ is the
 Hadamard transform, defined by $H\0{}=\0{}+\1{}$ and $H\1{}=\0{}-\1{}$.}.
 
 \begin{alltt}
1.  \(q \leftarrow i\)th qubit of \(W\sb{2,n-2}=\perm{\ket{1}\sp{\ox2}\ket{0}\sp{\ox(n-2)}}\)
    b = 0
    result = wait
2.  b := measure \(q\)
3.  if b = 1 then result := candidate
             else  result := voter 
4.  if result = voter then b := measure \(H(q)\);
                                broadcast b
\end{alltt}
Because $n$ is odd one of the candidates always gets more votes.  However,
in order to be able to appoint a leader, either the voters have to be able
to name the candidate they voted for, or the candidates must differ in that
they know which votes are intended for them.  Both possibilities violate
anonymity.  Note however, that the above protocol \emph{would} work when adapted
for a network where the communication graph is a ring and the processors
have a sense of direction.  Indeed, suppose each candidate sends a message
in say, the clockwise direction, such that the first candidate receiving
this message proclaims itself the leader.  This works because when $n$ is
odd both messages are ensured to arrive in different rounds.  Both time and
message complexity are in this case $\cO(n)$. 

\section{Quantum distributed consensus}\label{qdcsec}

\subsection{One qubit per processor}
\noindent
The results in this section are similar in spirit to those for leader election, in that they also depend on a symmetry property. However, this time it is symmetry preservation
rather than symmetry breaking which lies at the heart of the argument.  One caveat is that we are
considering a purely quantum solution to in this case. This is because with
classical post-processing there are more ways to break the symmetry
such that consensus is achieved, the most straightforward being to elect a
leader and have the leader distribute a value to all other processors.  Note
however, that this clearly would not work in a fault-tolerant setting.  In
summary, we find the following result.  

\begin{theorem}
A necessary and sufficient condition for a totally correct anonymous purely
quantum distributed consensus (QDC) protocol, where each processor owns 1 qubit
initially, is that processor qubits are entangled in a GHZ-state. 
\label{main2}
\end{theorem}
We prove this theorem in a similar manner as in \se{qleone}.

\paragraph{GHZ is sufficient.}The trick is to share a specific entangled state between all parties, which
allows one to create symmetrical knowledge in one step.  The state used is
known as the GHZ-state, where 

\begin{equation}
GHZ_{n}=\0{}^{\otimes n} + \1{}^{\otimes n}
\end{equation}
This state can be used as a symmetry-creating quantum resource.  Each
processor $i$ carries out the following protocol below. 
\begin{alltt}
1.  \(q \leftarrow i\)th qubit of \(GHZ\sb{n}\)
    result = wait
2.  result := measure \(q\)
\end{alltt}
This is a totally correct protocol with time complexity $\cO(1)$; no
message passing is required.  Note that the QDC protocol works as well for
different communication graphs or in asynchronous networks, and, more
importantly, in case an arbitrary number of processors is malicious.  This
is in contrast with the classical case where an all protocols are resilient
up to maximally $n/3$ faulty processors~\cite{Tel94}.

\paragraph{GHZ is necessary.} Suppose $P$ is an arbitrary protocol solving QDC. Then if $P$ allows $k$-symmetric paths with $k$ different $n$ it is not totally correct. Indeed, any $k$-symmetric path with $k < n$ results in a non-zero probability that only $k$ processors obtain symmetrical knowledge after the execution of $P$.. Thus neither partial correctness nor termination can be guaranteed, since
either $k$ processors terminate with different knowledge as the $n-k$
others, or they do not terminate precisely because of this.  Hence by
\pro{branch}, the initial network state $\gket$ should allow only
$n$-symmetric moves.  With $\gket \in (\comp^{2})^{\ox n}$ anonymous in the
sense of \de{anon}, this leaves us with $GHZ_{n}$, or unitary
transforms thereof, as the only possibility.

\subsection{More qubits per processor}
\noindent
Again one can repeat the same symmetry argument in the case where each
processor has $m$ qubits.  Taking $\{\ket{\phi_{i}}\}$ to be a basis for
$\cH$, this leads to initial network states of the form 

\begin{equation}
\gket=\sum_{i=1}^{2^{m}}\alpha_{i} \ket{\phi_{i}}^{\ox n}
\label{genGHZ}
\end{equation}
An $2^{m}$-valued QDC protocol would then consist of measuring this state.
Again, requiring total correctness means that one cannot but use states of
this type as a resource.

\section{Conclusion}\label{conclusion}
\noindent
The main contribution of this paper is a demonstration of the special
computational power of the W-state, and also of the GHZ-state, and
generalizations thereof.  A number of new results are established.  First,
totally correct leader election is trivially possible in anonymous quantum
networks; by itself this is not a major point.  Next, we prove that the
specific entanglement provided by the W- and GHZ-states, and their
generalizations, is the \emph{only} kind that exactly solves leader
election and purely quantum distributed consensus respectively.  The
W-state has been thought about less in quantum information theory than many
other entangled states though it does possess remarkable
properties~\cite{Dur00}.  It is highly persistent~\cite{Briegel01} for example, unlike
GHZ-states it requires many more measurements on average to destroy the entanglement.

In the programming languages community the relative power of synchronous vs.
asynchronous process calculi were compared using symmetry breaking
arguments: in fact on the ability to implement leader
election~\cite{Palamidessi03}.  The results of the present paper would have
similar consequences on the expressive power of quantum process calculi.
In joint work with others we are developing such calculi.

One can and should study the role of the W-state more thoroughly.  For
example what can be done with  a variant of the one-way model~\cite{Raussendorf01} based on
W-states? In this context, note that there is a difference between transformational
computing, where one has some inputs and is interested in computing some
output, and reactive computing, where the point of the algorithm is to
implement a behavior.  If one takes a distributed system where the
individual agents share a W-state and represents this as a classical data
structure one can simulate this efficiently classically. It is quite a different story to ask that a classical distributed system exhibit the same behavior as a distributed system sharing an entangled
quantum state.  As we have seen, there is no way that
a classical system can elect a leader.  We are not saying that with the W-state we can write transformational quantum programs that cannot be
simulated classically; we are saying that there are reactive behaviors
that are absolutely impossible classically.

There are possibly other quantum distributed algorithms using the W- state
that cannot be emulated at all (let alone efficiently) using classical
resources.  Perhaps new fault-tolerant behaviors can be realized.
We are actively investigating these and related questions.

\nonumsection{Acknowledgments}
\noindent
We would like to thank Elham Kashefi, Vincent Danos and especially Patrick
Hayden for helpful discussions.  This work was begun at the Oxford
University Computing Laboratory; we would like to thank that institution
for its hospitality.  Prakash Panangaden was supported by EPSRC (U.K.) and
NSERC (Canada) and Ellie D'Hondt by the Vrije Universiteit Brussel (Belgium).

\nonumsection{References}

\end{document}